\documentclass[fleqn,usenatbib,useAMS]{mnras}

\usepackage{newtxtext,newtxmath}
\usepackage[T1]{fontenc}
\DeclareRobustCommand{\VAN}[3]{#2}
\let\VANthebibliography\thebibliography
\def\thebibliography{\DeclareRobustCommand{\VAN}[3]{##3}\VANthebibliography}

\usepackage{graphicx}	
\usepackage{amsmath}	
\usepackage{multicol}        
\usepackage{bm}		
\usepackage{pdflscape}	
\usepackage[table]{xcolor}
\usepackage{xspace}

\newcommand{\shame}{{{\rm SHAM}e}\xspace}
\newcommand{\vpeak}{V_{\rm peak}}
\newcommand{\vmax}{V_{\rm max}}
\newcommand{\mpeak}{M_{\rm peak}}
\newcommand{\minfall}{M_{\rm infall}}

\newcommand{\hMsun}{ h^{-1}{\rm M_{ \odot}}}
\newcommand{\hMpc}{ h^{-1}{\rm Mpc}}

\newcommand{\ihMpcC}{ h^{3}{\rm Mpc}^{-3}}

\newcommand{\ihMpc}{ h\,{\rm Mpc}^{-1}}

\newcommand{\sig}{\sigma_{8}}
\newcommand{\OmM}{\Omega_\mathrm{M}}
\newcommand{\Omb}{\Omega_{\rm b}}

\newcommand{\ns}{{n_{\rm s}}}

\newcommand{\sigL}{\sigma_{\rm lum}}
\newcommand{\tmerger}{t_{\rm merger}}

\newcommand{\FkP}{f_{\rm k,cen+sat}}
\newcommand{\FkM}{f_{\rm k,cen-sat}}
\newcommand{\betaL}{\beta_{\rm lum}}
 
\newcommand{\proj}{${w_{\rm p}}$}
\newcommand{\mono}{$\xi_{\ell=0}$}
\newcommand{\quadr}{$\xi_{\ell=2}$}

\newcommand{\lensing}{$\Delta \Sigma$}

\newcommand{\Mr}{$M_{\rm r}$}

\usepackage[T1]{fontenc}
\usepackage{ae,aecompl}

\usepackage{newtxtext,newtxmath}

\title[Clustering and gg-lensing with SHAMe]{Consistent and simultaneous modelling of galaxy clustering and galaxy-galaxy lensing with Subhalo Abundance Matching}

\author[S. Contreras et al.]{
Sergio Contreras,$^{1}$\thanks{E-mail: \href{mailto:reangulo@dipc.org}{sergio.contreras@dipc.org}}
Raul E. Angulo$^{1,2}$\thanks{E-mail: \href{mailto:reangulo@dipc.org}{reangulo@dipc.org}},
Jon\'{a}s Chaves-Montero$^{1,3}$,
Simon D. M. White$^{4}$ and
Giovanni Aric{\`o}$^{1,5}$
\\
$^{1}$ Donostia International Physics Center, Manuel Lardizabal Ibilbidea, 4, 20018 Donostia, Gipuzkoa, Spain\\
$^{2}$ IKERBASQUE, Basque Foundation for Science, 48013, Bilbao, Spain.\\
$^{3}$ Institut de F\'isica d'Altes Energies, The Barcelona Institute of Science and Technology, Campus UAB, E-08193 Bellaterra (Barcelona), Spain.\\
$^{4}$ Max-Planck-Institut f\"{u}r Astrophysik, Karl-Schwarzschild-Str. 1, 85748, Garching, Germany\\
$^{5}$ Institute for Computational Science, University of Zurich, Winterthurerstrasse 190, 8057 Zurich, Switzerland.\\
}

\date{Accepted XXX. Received YYY; in original form ZZZ}

\pubyear{2020}

\begin{document}
\label{firstpage}
\pagerange{\pageref{firstpage}--\pageref{lastpage}}
\maketitle

\begin{abstract}
The spatial distribution of galaxies and their gravitational lensing signal offer complementary tests of galaxy formation physics and cosmology. However, their synergy can only be fully exploited if both probes are modelled accurately and consistently. In this paper, we demonstrate that this can be achieved using an extension of Subhalo Abundance Matching, dubbed \shame. Specifically, we use mock catalogues built from the TNG300 hydrodynamical simulation to show that  \shame can simultaneously model the multipoles of the redshift-space galaxy correlation function and galaxy-galaxy lensing, without noticeable bias within the statistical sampling uncertainties of a SDSS volume and on scales $r \in [0.6-30]\hMpc$. Modelling the baryonic processes in galaxy-galaxy lensing with a baryonification scheme allows \shame's range of validity to be extended to $r \in [0.1-30]\hMpc$. Remarkably, our model achieves this level of precision with just five free parameters beyond those describing the baryonification model. At fixed cosmology, we find that galaxy-galaxy lensing provides a general consistency test but little additional information on galaxy modelling parameters beyond that encoded in the redshift-space multipoles. It does, however, improve constraints if only the projected correlation function is available, as in surveys with only photometric redshifts. We expect \shame to have a higher fidelity across a wider range of scales than more traditional methods such as Halo Occupation Distribution modelling. Thus it should provide a significantly more powerful and more robust tool for analysing next-generation large-scale surveys.
\end{abstract}


\section{Introduction}
\label{sec:intro}

Modern cosmological observations can probe multiple aspects of the large-scale structure (LSS) of the Universe: the distribution of galaxies via its clustering statistics, the distribution of mass via weak gravitational lensing, and the distribution of ionized gas via Sunyaev-Zeldovich effects. Although each probe can be analysed independently, gas, dark matter, and galaxies are inherently linked via physics and cosmology. Thus, simultaneous analysis of multiple probes offers a unique opportunity to learn about both galaxy formation physics and the underlying cosmological model. 

A prime example of complementarity among cosmological observables is the case of galaxy clustering and weak gravitational lensing. According to our current understanding of galaxy formation, galaxies form in dark matter halos and their properties are shaped by halo evolution. Therefore, galaxy clustering (GC) is in part determined by the location and velocity of dark matter halos and subhalos. At the same time, the gravitational lensing around galaxies, known as galaxy-galaxy lensing (GGL), is dominated by the correlation between galaxy positions and the underlying mass distribution. Therefore, a joint analysis of clustering and lensing can inform us about the galaxy-halo connection, can break degeneracies between galaxy formation physics and cosmology, and can ultimately improve our understanding of cosmic structure. 

Nowadays, clustering and lensing are routinely analysed jointly by LSS surveys, improving considerably the cosmological parameter constraints extracted using each probe alone. For instance, the Dark Energy Survey (DES) collaboration finds $\sim25\%-30\%$ stronger constraints on $S_8\equiv \sig (\OmM/0.3)^{0.5}$ when galaxy clustering and GGL are included in addition to cosmic shear correlations \citep{Secco:2022,Amon:2022}. Similarly, the improvement for the KIDS-1000 survey is about $30\%$ on $S_8$ \citep{Asgari:2021,Heymans:2021}. Reflecting this, a joint lensing and clustering analysis forms a key part of the science analysis plan for upcoming surveys such as Rubin's LSST.

Naturally, a key ingredient for such analyses is a theoretical model for the signal that should at least match the statistical accuracy of the observations. The most common approach adopted in LSS analyses is a perturbative description of galaxy bias \citep[see][for a review]{Desjacques:2018}. This approach has the advantage of being general and so should be applicable to any kind of galaxy, regardless of galaxy formation physics and/or selection criteria. On the other hand, this perturbative approach is only valid on relatively large scales. For instance, \cite{Sugiyama:2020} showed that a simple linear bias model is valid down to $8$ and $12$ $\hMpc$, for the projected correlation function and GGL, respectively, which are then lower limits on the scales that should be used in cosmological analyses \citep[e.g.][]{Krause:2021,Sugiyama:2022}. Recently, hybrid approaches have emerged as an alternative to extend somewhat the range of validity of a perturbative bias expansion \citep{Zennaro:2021,Kokron:2021,Zennaro:2022,Pellejero-Ibanez:2022}, allowing the extraction of more information from current datasets \citep{Hadzhiyska:2021}. 

Additional modelling is required to access even smaller, non-perturvative, scales. The most common approach is the so-called Halo Occupation Distribution model (HOD). This assumes a functional form for the abundance of galaxies in halos of a given mass, which, in combination with $N$-body simulations, can then provide clustering and GGL  predictions down to arbitrarily small scales \citep[see e.g.][]{vandenBosch:2013}. An HOD-based approach was employed by \cite{Cacciato:2013}, \cite{More:2015} and \cite{Miyatake:2022} to analyse SDSS clustering and GGL, obtaining competitive cosmological parameter constraints. In addition, \cite{Dvornik:2022} adopted a similar approach and re-analysed GGL and GC measurements from KIDS-1000, obtaining constraints on $S_8$ similar to those obtained by a standard 3x2pt analysis -- this illustrates the potential for obtaining cosmological information from data currently discarded by most cosmological analyses. However, it is important to bear in mind that HOD modelling makes strong assumptions about the galaxy-halo connection, some of which have been shown to be incorrect in our best current theories of galaxy formation \citep{Croton:2007,ChavesMontero:2022}. This could create apparent discrepancies between GC and GGL and potentially bias cosmological parameter estimates.

The most sophisticated methods for making joint predictions for GC and GGL are Semi-Analytic Models of galaxy formation  (SAMs) and hydrodynamical simulations. By modelling directly the main physical processes that govern nonlinear gravitational evolution and galaxy formation, the connection between GC and GGL emerges naturally. Although these have not so far been employed for cosmological inference (due to their computational cost), their predictions can be compared directly to observation in order to test and enhance the physical recipes they implement. For instance, \cite{Renneby:2020} compared the predictions of the TNG300 hydrodynamical simulation to observations from KIDS, showing them to be in relatively good agreement. On the other hand, \cite{Wang:2016} studied GGL and GC as measured in SDSS around locally-brightest galaxies, finding significant tension with the predictions of the \cite{Henriques:2015} version of the L-Galaxies SAM. The \cite{Guo:2011} version was somewhat more successful, a conclusion that was extended to subsamples split by galaxy colour by \cite{Mandelbaum:2016}. More recently, \cite{Renneby:2020} employed GGL data to constrain the free parameters of the L-Galaxies model, finding that shorter dynamical friction timescales were needed to improve the agreement with observations for galaxies with stellar masses above $10^{11}\hMsun$. Although no modification was found adequate to describe the GGL signal around less massive galaxies, this served as an example of how a joint GGL and GC analysis can be used to study the underlying physical processes.

As argued above, HOD methods may not be accurate enough to analyse current datasets and SAM/Hydro methods are computationally too expensive to sample a large high-dimensional cosmological+galaxy formation parameter space. An alternative is Subhalo Abundance Matching (SHAM \citealt{Vale:2006, Conroy:2006}), which may provide a more realistic description of the (sub)halo-galaxy connection than HODs with more flexibility and computational speed than models that seek to directly follow physical processes. Nonetheless, the standard SHAM approach has also found it difficult to reproduce GC and GGL simultaneously in observational samples (eg. \citealt{Leauthaud:2017}, see also \citealt{RodriguezTorres:2016,Saito:2016} for a full description of the SHAM models used in that work)

This paper is the fourth in a series where we have developed a flavour of Subhalo Abundance Matching that has enough flexibility and accuracy to interpret modern large-scale structure surveys. In \cite{C21a} we proposed to use the large-scale environment of each dark matter subhalo to model so-called galaxy assembly bias. Then, in \cite{C21c} we use this property, together with modelling of orphan subhalos and tidal disruption, to improve upon classical SHAM models. Explicitly, we showed that it accurately fits redshift-space galaxy clustering as measured in various semi-analytical galaxy formation models and in the TNG300 hydrodynamical simulation. In \cite{C22a} we combined \shame with a $N$-body rescaling algorithm to model galaxy clustering as a function of cosmology, and showed it can successfully recover unbiased cosmological parameters in the state-of-the-art Millennium-TNG simulation \citep{Bose:2022, Kannan:2022, Hadzhiyska:2022b,Hadzhiyska:2022a, Barrera:2022, Pakmor:2022, Hernandez:2022}. 

In this paper, we investigate whether \shame can simultaneously describe GC and GGL. For this purpose, we will employ the TNG300 simulation to build 3 different galaxy catalogues for which we measure the projected correlation function, the monopole and quadrupole of the redshift-space correlation function, and the average surface mass density. As mentioned above, TNG300 galaxies have been shown to be in reasonably good agreement with observed GC and GGL, thus they represent a realistic benchmark to test our model. We will show that \shame can indeed describe all these statistics simultaneously to better accuracy than the statistical uncertainty of the TNG300 results. Importantly, this is not only true for this specific simulation, but SHAMe displays the same performance when applied to 4 different SAM catalogues with extreme values of their free parameters. The results of this paper validate further the physical soundness and robustness of \shame, and therefore lend support to its use for extracting cosmological constraints from GC and GGL.

This paper is organised as follows. In \S\ref{sec:simulations} we describe our mock galaxy catalogues extracted from the TNG simulations. We also briefly summarise the main aspects of \shame. In \S\ref{sec:methodology} we describe the main numerical methods we employ, including details on how we compute mock galaxy-galaxy lensing measurements, the way in which we estimate covariance matrices, and how we build an emulator for clustering and lensing. In \S\ref{sec:cluster} we examine the performance of SHAMe for describing all statistics, and in \S{\ref{sec:baryons}} we focus on the so-called baryonic effects. We continue in \S\ref{sec:other_stat} by presenting other statistics not used in calibrating the free parameters of \shame. We finalise in \S\ref{sec:summary} presenting our conclusions and summarising our main findings. 

\section{Numerical simulations \& galaxy population models}
\label{sec:simulations}

In this section, we describe the TNG300 hydrodynamic simulation (\S~\ref{sec:tng}) and a suite of gravity-only simulations we employed (\S~\ref{sec:DMonly}), as well as our ``Subhalo Abundance Matching extended'' model -- \shame (\S~\ref{sec:shame}).

\subsection{The TNG300}
\label{sec:tng}

To validate \shame, we use galaxy samples from the Illustris-TNG300 simulation (thereafter TNG300). This simulation is part of ``The Next Generation'' Illustris Simulation suite of magneto-hydrodynamic cosmological simulations (IllustrisTNG, \citealt{TNGa, TNGb, TNGc, TNGd, TNGe}), successors of the original Illustris simulation \citep{Illustrisa,Illustrisb,Illustrisc,Illustrisd}. The TNG300 is one of the largest publicly available high-resolution hydrodynamic simulations in the world\footnote{\url{https://www.tng-project.org/}}. The simulated volume is a periodic box of 205 $\hMpc$ ($\sim300$ Mpc) a side. The number of dark matter particles and gas cells is $2500^3$ each, implying a baryonic mass resolution of $7.44\times10^6\,\hMsun$ and a dark matter particle mass of $3.98\times 10^7\,\hMsun$. The simulations adopted cosmological parameters consistent with recent analyses \citep{Planck2015}\footnote{$\OmM$ = 0.3089, $\Omb$ = 0.0486, $\sig$ = 0.8159, $\ns$ = 0.9667 and $h$ = 0.6774.}.

The TNG suite of simulations was run using \texttt{AREPO} \citep{AREPO} and features a series of improvements upon their predecessor, the Illustris simulation, including i) an updated kinetic AGN feedback model for the low accretion state \citep{Weinberger:2017}; ii) an improved parameterization of galactic winds \citep{Pillepich:2018}; and iii) the inclusion of magnetic fields based on ideal magneto-hydrodynamics \citep{Pakmor:2011,Pakmor:2013,Pakmor:2014}. 

The free parameters of the model were calibrated so that the simulation agrees with several observations: the stellar mass function, the stellar-to-halo mass relation, the total gas mass content within the virial radius ($r_{500}$) of massive groups, the stellar mass – stellar size and the BH–galaxy mass relations all at z = 0, and the overall shape of the cosmic star formation rate density up to z $\sim$ 10. Nevertheless, the TNG results successfully reproduce many observables not directly employed in the calibration process (e.g. \citealt{TNGb, TNGd,Vogelsberger:2020}). Most important for our analysis, the clustering of TNG galaxies is in good agreement with observational estimates over a broad range of stellar masses \citep{Springel:2018}. In addition, as shown by \cite{Renneby:2020}, the TNG300 catalogues agree well with the KIDS+GAMA and SDSS observations for GGL, including for selection using stellar mass, colour, group membership, and isolation criteria. However, we would like to emphasise that we do not necessarily require the catalogues to be a perfect match to observations. Instead, we are interested in the ability of our model to describe the GC and GGL of multiple galaxy formation models that provide reasonable and physically-consistent predictions. In other words, we seek a model with enough flexibility to describe a broad class of realistic models, which could then increase our confidence that it can correctly describe the real Universe. 

We will test the GC and GGL predictions of \shame against 3 catalogues with different number densities at $z=0$. To build these samples, we define the luminosity of a galaxy as the sum of the luminosities of all the stellar particles in a group. We then choose the most luminous galaxies in the r-band with the following number densities: $10^{-2},\ 3.16\times10^{-3}\ \&\ 8.7\times10^{-4}\ \ihMpcC$, equivalent to a magnitude cut of -20.29, -21.28 and -22.02 and a median halo mass of $1.47\ 10^{12}\hMsun$, $3.24\ 10^{12}\hMsun$ and $1.02\ 10^{13}\hMsun$ respectively for the TNG300.

\subsection{The gravity-only simulations}
\label{sec:DMonly}

To ease the comparison of \shame with the TNG300, we carried out a gravity-only simulation with the same initial conditions as the TNG, but with a reduced resolution. We refer to this simulation as ``TNG300-mimic''. The simulation has $4^3$ fewer particles ($625^3$ particles) with a mass of $3.03\times 10^9 h^{-1}M_{\odot}$. It was carried out with an updated version of {\tt L-Gadget3} \citep{Angulo:2012}, a lean version of {\tt GADGET} \citep{Springel:2005} used to run the Millennium XXL simulation and the Bacco Simulations \citep{Angulo:2021}. This version of the code allows an on-the-fly identification of haloes and subhaloes using a Friend-of-Friend algorithm \citep[{\tt FOF}][]{Davis:1985} and an extended version of {\tt SUBFIND} \citep{Springel:2001}. Our updated version of {\tt SUBFIND} can better identify substructures by considering their past history, while also measuring properties that are non-local in time, such as the peak halo mass ($\mpeak$), peak maximum circular velocity ($\vpeak$), infall subhalo mass ($\minfall$), and mass accretion rate, among others. Our version of {\tt SUBFIND} also allowed us to identify orphan substructures, i.e.~satellite structures with known progenitors that the simulation cannot resolve but that are expected to exist in the halo; these are used by our empirical model. This simulation only took 945 CPU hours, where about 13\% of the time was spent in FOF/subfind calculations.

We note that we employ the same numerical accuracy parameters as the Bacco simulations \citep{Angulo:2021}. As validation, we have compared the $z=0$ matter power spectrum of our simulation to that measured in the TNG300-3-Dark, a public dark matter-only simulation with the same characteristic as the TNG300-mimic, finding differences below 0.8\% in the matter power spectrum at all scales.

For our estimation of covariance matrices, which we will discuss in \S\ref{sec:Cv}, we will use one of the Bacco simulations, ``The One'' scaled to the cosmology of the TNG300 following the procedure of (\citealt{Angulo:2010,C20}). The simulation has $4320^3$ particles and a box length of 1440 $\hMpc$ ($\sim1400 \hMpc$ after the cosmology scaling), from which we can extract 340 regions of the same volume as the TNG300. We note that this simulation was run with opposite initial Fourier phases, using the procedure of \cite{Angulo:2016} that suppresses cosmic variance on large scales by up to 50 times compared to a random simulation of the same volume. However, this procedure has little impact on the scales of interest in this study, thus the simulations can be regarded as if they evolved a fully random initial Gaussian field. 

\subsection{Sub-halo abundance matching extended model}
\label{sec:shame}

To extract information from GC and GGL, we require a model capable of realistically and efficiently populating dark matter simulations. To accomplish this, we employ the {\bf S}ub-{\bf H}alo {\bf A}bundance {\bf M}atching {\bf e}xtended model (\shame) developed by \cite{C21c}. The two primary advantages of this model are (a) the small number of free parameters compared to other models such as HODs and (b) the precision with which this model reproduces galaxy clustering in redshift space, particularly on small scales. In this paper, we employ the same version of \shame as in \cite{C22a}, which focuses on reproducing galaxy clustering using $\rm M_r$-selected samples. 

As in the standard SHAM model \citep[][]{Vale:2006,Conroy:2006,Reddick:2013,C15,ChavesMontero:2016,Lehmann:2017,Dragomir:2018,Hadzhiyska:2021b} the model begins by matching a subhalo property (in this case, $\vpeak$) to the expected luminosity function. We define $\vpeak$ as the maximum circular velocity ($\vmax \equiv {\rm max}(\sqrt{GM(<r)/r})$) during the subhalo evolution. We use the luminosity function of the TNG300 to match our galaxies. As mentioned in \cite{C21c}, when galaxies are selected using number density cuts, the choice of a specific luminosity function has little to no effect on the galaxy clustering statistics. From here, we apply three different prescriptions to improve the GC predictions: inclusion of orphan galaxies, disruption or decrement of the luminosity of satellite galaxies, and modelling of galaxy assembly bias in the galaxy sample.
 
\subsubsection*{Adding orphan galaxies}

Orphan galaxies are satellite structures with a known progenitor that the simulation cannot resolve, but that are nevertheless expected to survive in the dark matter halo. While they were originally introduced to account for the limited resolution of a dark matter simulation, \cite{Guo:2014} showed that orphan galaxies may be necessary for a SHAM model to reproduce the galaxy clustering on small scales even on reasonably-high resolution simulations.

We assume that an orphan galaxy merges with its central structure when the time since accretion ($t_{\rm infall}$) becomes larger than a dynamical friction timescale ($t_{\rm d.f.}$). We compute $t_{\rm d.f.}$ using a modified version of Eq. 7.26 of \cite{BT:1987} at the moment the subhalo becomes an orphan.

\begin{equation}
t_{\rm d.f.} = \dfrac{1.17\ \tmerger\ d_{\rm host}^2\ V_{\rm host} (M_{\rm host}/10^{13}\ h^{-1}\mathrm{M}_{\odot})^{1/2}}{G \ln(M_{\rm host}/M_{\rm sub}+1)\ M_{\rm sub}},
\end{equation}

\noindent where $\tmerger$ is a free non-dimensional parameter that effectively regulates the number of orphan galaxies; $\rm d_{host}$ is the distance of the subhalo to the centre of its host halo; $\rm v_{host}$ is the virial velocity of the host halo; $\rm M_{host}$ is the virial mass of the host halo; $\rm M_{sub}$ is the subhalo mass, and $G$ is the gravitational constant.

\subsubsection*{Luminosity attenuation}

After accretion, satellite galaxies progressively lose their gas due to ram pressure stripping, which causes their star formation rate to decrease. As a result, satellites stop forming newer blue, massive stars, and those in the galaxy successively leave the main sequence, which turns optical colours redder and decreases the luminosity in optical bands \citep[e.g.,][]{chaves:2021}. Eventually, satellite galaxies also lose their stars due to tidal forces, reducing their luminosity even further.

To account for this, we exclude all galaxies that have been satellites for an extended time, i.e. $t_{\rm infall} > \betaL\, t_{\rm dyn}$, with $t_{\rm dyn}$ the halo's dynamical time, defined as $0.1/H(z)\sim 1.44\,\rm{Gyrs}$ and $\betaL$ being a free parameter. We find that this simple approach can substantially improve the GC predictions for satellite galaxies. We also test alternative approaches, such as removing substructures that have lost a considerable amount of mass (as in \citealt{C21a} and \citealt{Moster:2018}) and other more complex ones, but find that our simple approach fits the galaxy clustering the best for a luminosity-selected galaxy sample.

\subsubsection*{Assembly bias}

Galaxy assembly bias \citep{Croton:2007} is caused by the propagation of halo assembly bias \citep{Gao:2005, Gao:2007} into galaxy clustering. This propagation occurs because the occupation of galaxies depends on halo properties other than mass, and these properties depend on large-scale environment through halo assembly bias (this effect is also known as occupancy variation; \citealt{Zehavi:2018, Artale:2018}). As far as we are aware, there has been no definitive confirmation of the (non)existence of this form of galaxy assembly bias in the real Universe.

The amount of assembly bias of a hydrodynamic simulation differs from that of a SHAM \citep{ChavesMontero:2016} or of SAMs \citep{C21a}. Furthermore, these levels do not need to correspond to the level of assembly bias in the real Universe. To account for the uncertainty surrounding the assembly bias level of the target galaxy sample, we modify our \shame procedures to emulate any possible assembly bias level. To do so, we employ the method developed by \cite{C21a} which uses the individual bias-per-object of the haloes \citep{Paranjape:2018} to bias the galaxy content of haloes according to their environment. In essence, the model swaps the luminosities of galaxies with similar values of $\vpeak$ so that their luminosities can correlate or anticorrelate with large-scale environment density. We preserve the satellite fraction of the original galaxy sample by performing this operation separately for central and satellite galaxies. This method uses two free parameters to control the level of bias of the galaxies, $f_{\rm k,cen}$ and $f_{\rm k,sat}$, for central and satellite galaxies, respectively. A value of $f_{\rm k}=1$ (-1) indicates a maximum (minimum) galaxy assembly bias signal, while a value of 0 means the same assembly bias level as a standard SHAM. 

In contrast to \cite{C22a}, we do not force  $f_{\rm k,cen}$ = $f_{\rm k,sat}$. They used this approximation to reduce the number of parameters since they found no reduction in the power to constrain cosmology when these parameters were set equal. Throughout this paper, we will express these parameters as $\FkP$ and $\FkM$, which respectively represent $(f_{\rm k,cen}$+$f_{\rm k,sat})/2$ and $(f_{\rm k,cen}$-$f_{\rm k,sat})/2$. 
\subsection{Semi-analytical model}
\label{sec:sam}

The TNG300 simulation is arguably one of the most realistic galaxy formation simulations available. However, it is still an approximation to the real Universe, therefore it is important to explore whether \shame can also accurately describe GC and GGL in other galaxy formation models. To address this question, we have carried out semi-analytical modelling of galaxy formation (SAM, e.g. \citealt{Kauffmann:1993, Cole:1994, Bower:2006, Lagos:2008, Benson:2010, Benson:2012, Jiang:2014, Croton:2016, Lagos:2018, Stevens:2018, Henriques:2020}). We use the public version of \texttt{L-Galaxies}\footnote{\url{http://galformod.mpa-garching.mpg.de/public/LGalaxies/}}, the SAM developed by the ``Munich group'' \citep{White:1991, Kauffmann:1993, Kauffmann:1999, Springel:2001,DeLucia:2004, Croton:2006, DeLucia:2007, Guo:2011, Guo:2013, Henriques:2013,Henriques:2020} and is specifically based on the model of \cite{Henriques:2015}. We run \texttt{L-Galaxies} five times on the TNG300-mimic dark matter simulation. One time using (mostly) its default parameter set, while the other four were run with the supernova energy efficiency parameter multiplied/divided by a factor of 10 or 100. We opted to change this parameter because it was the one that maximally influenced the clustering of \Mr~selected galaxy samples at a fixed number density. 

\section{Methodology}
\label{sec:methodology}

In this section we will provide details on how we compute the GC and the GGL signal in our catalogues (\S\ref{sec:lensing}), and the respective covariance matrix (\S\ref{sec:Cv}). We also describe the construction of an emulator for GC and GGL (\S\ref{sec:emu}) with which we will later fit the TNG and SAM mock catalogues.

\subsection{Clustering \& lensing}
\label{sec:lensing}

\subsubsection*{Galaxy clustering}

The GC statistics we used in this work are the projected correlation function (\proj) and the multipoles of the correlation function, specifically, the monopole (\mono) and the quadrupole (\quadr). We utilised \textsc{corrfunc} \citep{Corrfunc1,Corrfunc2} to compute the projected correlation function, which is obtained by integrating the 2-point correlation function, $\xi(\rm p, \pi)$, over the line of sight, 
\begin{equation}
    w_{\rm p} = 2 \times \int_{0}^{\pi_{\rm max}} \xi(\rm p, \pi)d\pi
\end{equation}
\noindent with $\pi_{\rm max}=30 \hMpc$ the maximum deep used on the pair counting. We do not consider larger scales because the results become too noisy due to the small size of the TNG300 simulation box.

To measure the multipoles correlation function, we first measure the 2-point correlation function in bins of $s$ and $\mu$, where $s^2 = r^2_{\rm p} + r^2_{\pi}$ and $\mu$ is the cosine of the angle between $s$ and the direction of the line-of-sight. We again compute this statistic using \textsc{corrfunc}. The multipoles correlation functions are then defined as, 
\begin{equation}
    \xi_{\ell} = \frac{2 \ell+1}{2} \int^{1}_{-1} \xi(s,\mu)P_\ell(\mu)d\mu
\end{equation}
\noindent with $P_{\ell}$ is the $\ell$ -th order Legendre polynomial.

\subsubsection*{Galaxy-galaxy lensing}

A foreground mass distribution induces a shear signal on background sources that depends upon the distances between the lens-source pair. The stacked signal from a sufficiently large number of lenses is proportional to the excess surface density
\begin{equation}
    \Delta\Sigma(r_\perp) = \overline{\Sigma}(\leq r_\perp) - \Sigma(r_\perp),
\end{equation}
where $\Sigma$ is the azimuthally-averaged surface mass density and $\overline{\Sigma}(\leq r_\perp)$ is the mean surface density within projected radius $r_\perp$. We estimate the azimuthally-averaged surface mass density using
\begin{equation}
    \Sigma(r_\perp) = \Omega_\mathrm{m} \rho_\mathrm{crit} 
    \int_{-r_\parallel^\mathrm{max}}^{r_\parallel^\mathrm{max}} \xi_\mathrm{gm}(r_\perp, r_\parallel)\,\mathrm{d}r_\parallel,
\end{equation}
where $r_\parallel$ and $r_\perp$ refer to the projected distance along and perpendicular to the line-of-sight from the lens galaxy, $r_\parallel^\mathrm{max}$ is the integration boundary, $\xi_\mathrm{gm}$ is the galaxy-matter three-dimensional cross-correlation, $\Omega_\mathrm{m}$ and $\rho_\mathrm{crit}$ are the matter and critical density of the Universe, respectively, and the mean surface mass density within a radius $r_\perp$ is
\begin{equation}
    \overline{\Sigma}(\leq r_\perp) = \frac{2}{r_\perp^2}\int_0^{r_\perp} \Sigma(\tilde{r})\,\tilde{r} \,\mathrm{d}\tilde{r}.
\end{equation}

We measure $\xi_{\rm gm}$ using \textsc{corrfunc}, $r_\parallel^\mathrm{max}=30\hMpc$, and a subsampled version of matter density field diluted by a factor of $\sim 1/3000$. We checked that this dilution factor ensures $\Delta\Sigma$ measurements with subpercent precision across the entire range of scales considered in this work. To reduce the uncertainty in our calculations, we compute the average of each observable after considering the line-of-sight to be each of the three simulation coordinate axes (the same is done for all GC statistics).

\subsection{Covariance matrices}
\label{sec:Cv}

To quantify the accuracy of the model, we build the covariance matrices based on the expected errors from an SDSS-like survey. To do this, we scaled one of the Ring Simulations, ``The One'', to the cosmology of the TNG300 using the ``scaling technique'' described in (\citealt{Angulo:2012}, see also \citealt{C20}). This is to account for the differences in $\sig$, the only cosmological parameter that differs significantly between simulations. In a nutshell, this method shifts the particles, haloes, and subhaloes of a simulation of dark matter so that its mass distribution resembles that of a simulation with a different cosmology. The box length of this scaled simulation is $\sim 1400 \hMpc$. We divided this volume into 64 smaller boxes of $(350 \hMpc)^3$, a volume comparable to that of the SDSS galaxy sample for a \Mr=$-20.5$ sample, which has a number density of $3.13\ \ihMpcC$, similar to our middle dense sample \citep{Guo:2015}. ``The One'' was not run with gaussian initial conditions, but rather with the fixed-and-paired technique developed by \cite{Angulo:2016}, which reduces the effect of cosmic variance on a simulation by a significant amount. While not gaussian at large scales, we find that the GC and GGL signals at the scales we are interested ($<25 \hMpc$) are mostly gaussian, meaning they can be used to compute our covariance matrix.
 
We used a \shame model to populate each of the subboxes, which uses the parameters obtained by fitting simultaneously the TNG300 GC and GGL for each of the different luminosity samples at scales below 0.6 $\hMpc$ (assuming jackknife errors on the TNG300 data, eg. \citealt{Zehavi:2002,Norberg:2008}).

We compute the covariance matrix ($C_v$) as:

\begin{equation}
C_v(V_i,V_j) = \frac{1}{N} \sum_{k=1}^{N}(V_i^k-\bar{V_i})(V_j^k-\bar{V_j}),
\end{equation}

\noindent where $N$ is the number of samples (in this case 64) and $V$ is a concatenation vector of our summary observables, $V$=\proj, \mono, \quadr, \& \lensing. The use of a single vector containing all of the GC and GGL information implies that the covariance matrix will account for the cross-covariance between the different statistics. To avoid problems with the non-periodic boundary conditions of our subvolumes, the GC statistics are computed using the cross-correlation between the galaxies of the subvolume and the galaxies in the entire box and for the GGL use the same dark matter particles from the full simulation. To account for the need for the SHAMe model to fit different galaxy formation models \citep{C22a} and other possible systematics (like the error of the emulator or shot noise), we add an additional 5\% of the clustering signal to the diagonal of the covariance matrices. We also test adding 3\% instead, finding similar results as with 5\%.

To test the quality of the inversion of the covariance matrix, we look at the non-diagonal terms of $\rm C_v\ C_v^{-1}$), finding that all terms are below $10^{-9}$. 
 
\subsection{Emulator}
\label{sec:emu}

To speed up our analysis, we built an emulator for GC and GGL using $\sim 121.000$ mock  measurements with varying \shame parameter values randomly. The range of the \shame parameters we look at are:

\begin{eqnarray}
\label{eq:par_range}
\sigL                      &\in& [0,    0.3]\\
\log(\tmerger)              &\in& [-1.5, 1.2]\\
\FkP                       &\in& [-0.5, 0.5]\\
\FkM                       &\in& [-0.5, 0.5]\\
\betaL                     &\in& [ 2.5,  10.5]
\end{eqnarray}
\noindent where $\sigL$ is the scatter between $\vpeak$ and \Mr;  $\tmerger$ is a dimensionless parameter that effectively regulates the number of orphan galaxies; $\FkP$ and $\FkM$ are the assembly bias parameters and $\betaL$ is a parameter that regulates the number of non-orphan satellite galaxies. We refer to \S\ref{sec:shame} for further details about these parameters.

Similar to \cite{Angulo:2021, Arico:2021a,Arico:2021b,Pellejero:2022b, Zennaro:2021b}, we construct our emulators using a feed-forward Neural Network. The architecture consists of two fully-connected hidden layers with 200 neurons each and a Rectified Linear Unit activation function for the projected correlation function and the monopole of the correlation function, as well as three layers for the quadrupole and the lensing, with each statistic represented by a separate network. We explored alternative configurations obtaining comparable performances.

We built the neural networks using the Keras front-end of the Tensor-flow library. We used the Adam optimization algorithm with a learning rate of 0.001 and a mean squared error loss function. Our dataset is divided into two separate groups: 90\% of the data was used for training and the remaining for validation. On a single Nvidia Quadro RTX 8000 GPU card, the training required approximately 20 minutes per number density/statistic. Evaluating the four emulators on a personal laptop takes $\sim$ 0.1 seconds while evaluating 100,000 samples takes $\sim$ 1 seconds on a laptop (it is more efficient to evaluate the emulator in large batches). Depending on the statistic and the number density, the precision of the emulator ranges from 0.5\% to 3\%.

\section{Galaxy clustering and galaxy-galaxy lensing in \shame}
\label{sec:cluster}

\begin{figure*}
\includegraphics[width=0.95\textwidth]{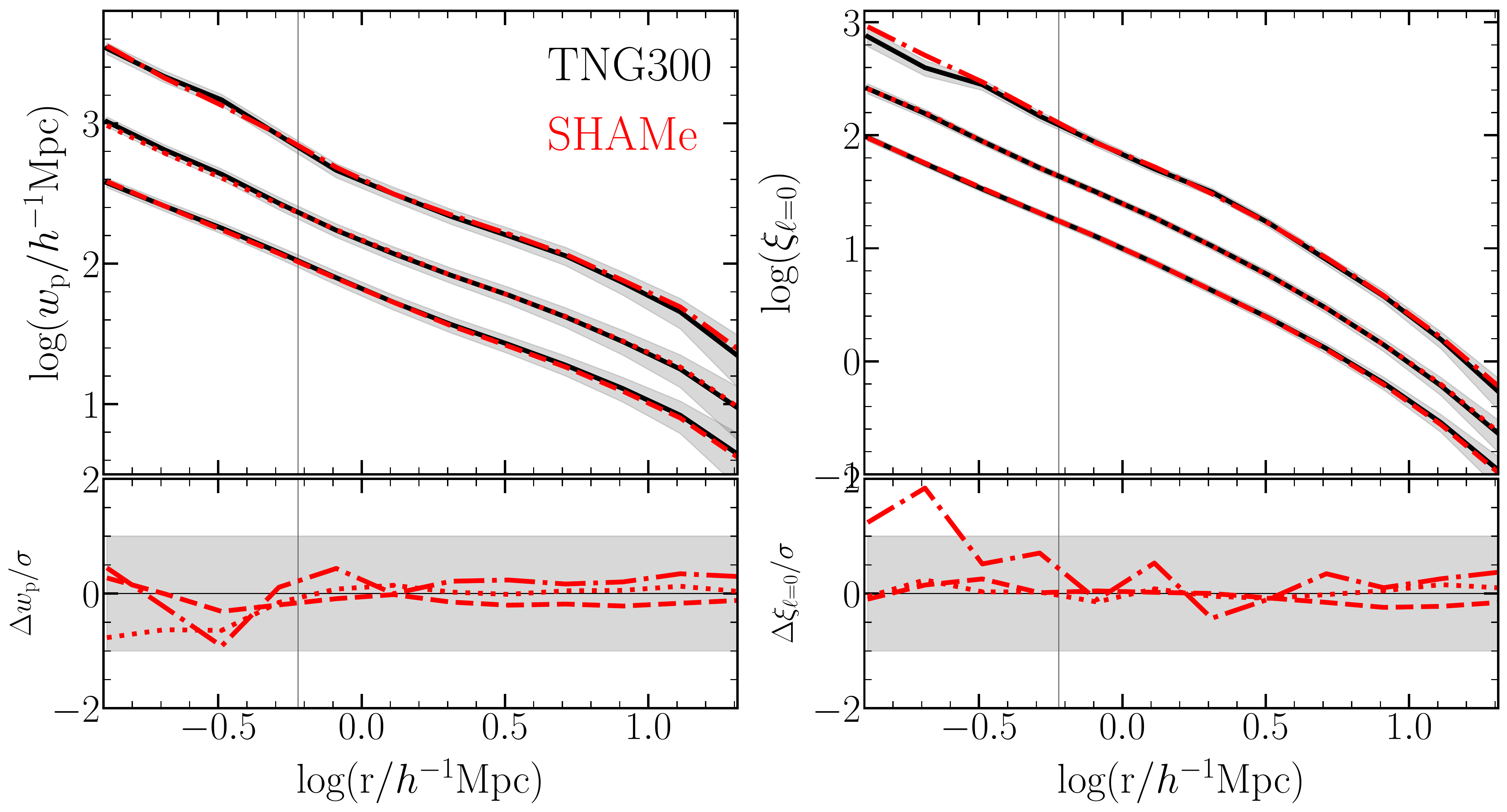}
\includegraphics[width=0.95\textwidth]{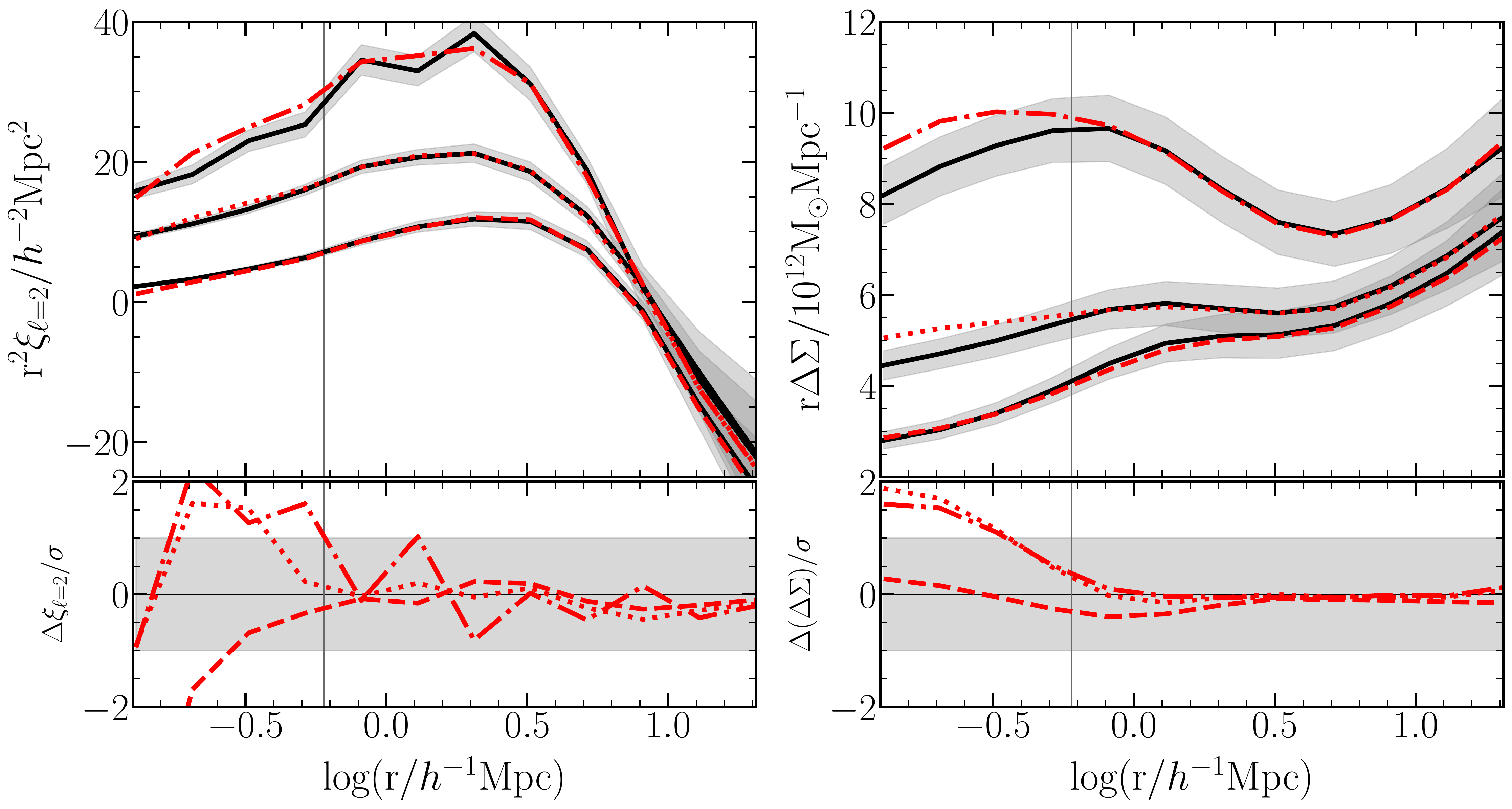}
\caption{The TNG300 simulation's (black lines) and the SHAMe model's (red lines) galaxy clustering and galaxy-galaxy lensing. The upper left, upper right, lower left, and lower right panels represent the projected correlation function, monopole, quadrupole, and lensing, respectively. The bottom, middle, and top lines in each panel show the results for samples with a number density of $10^{-2},\ 3.16\times10^{-3}\ \&\ 8.7\times10^{-4}\ \ihMpcC$, respectively. For clarity, the lines in the projected correlation function, monopole, and quadrupole have been shifted along the y-axis by 20 and 40 for the $3.16\times10^{-3}\ \&\ 8.7\times10^{-4}\ \ihMpcC$ samples. The vertical line at 0.6 $\hMpc$ represents the SHAMe model's minimum scale for fitting the TNG300 clustering. The TNG300's shaded region represents the square root of the diagonal elements of the covariance matrix used in the fitting. In each panel, the bottom plot shows the difference between TNG300 and SHAMe clustering divided by the square root of the diagonal term of the covariance matrix used for fitting.}  
\label{Fig:gal_cluster}
\end{figure*}

\begin{figure*}
\includegraphics[width=1.0\textwidth]{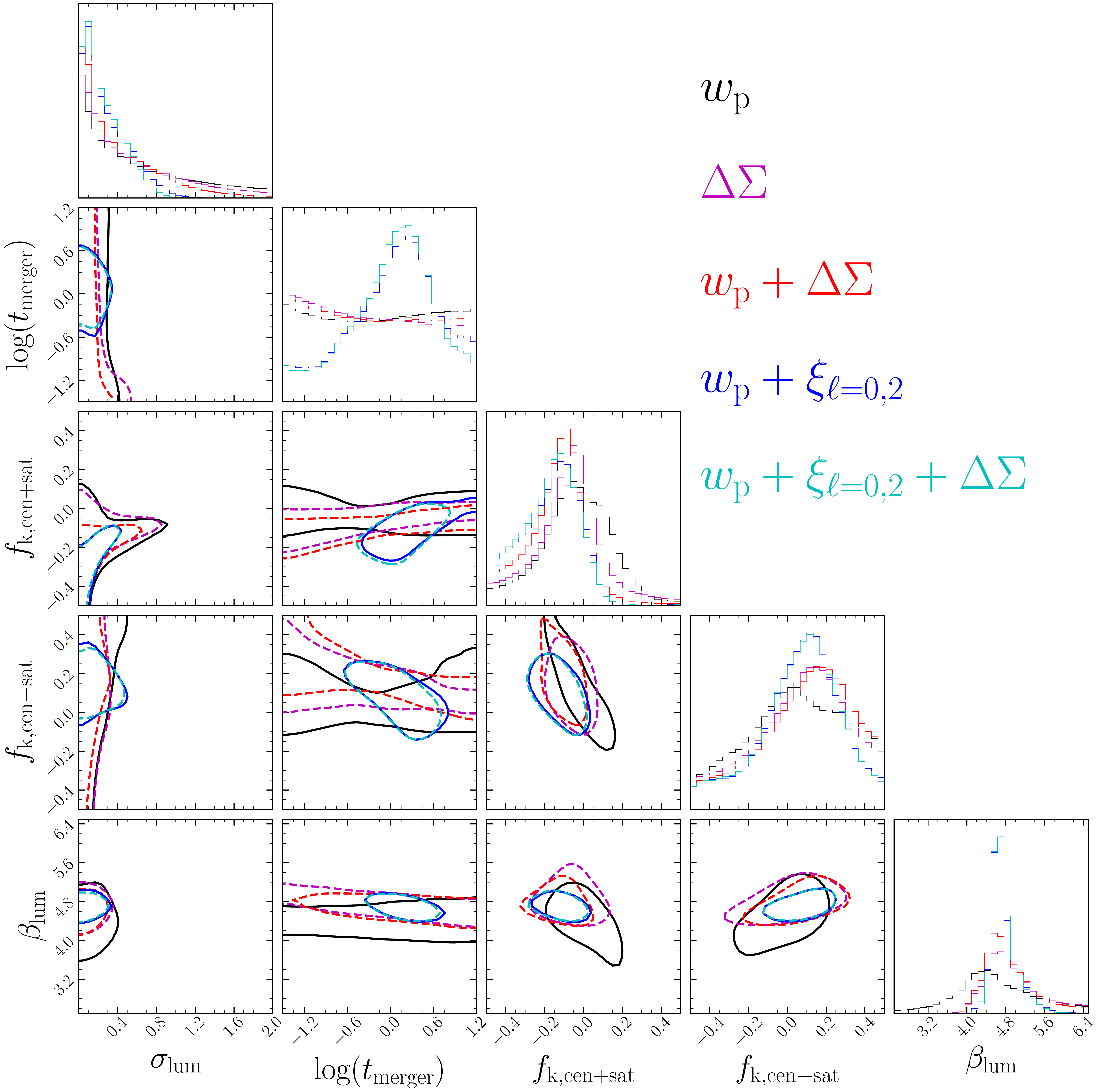}
\caption{Marginalised 1$\sigma$ credibility regions for five SHAMe parameters derived from the clustering (\proj,~\mono,~\quadr)~and~lensing ($\Delta \Sigma$) of the TNG300 simulation at a number density of n=$0.00316\ \ihMpcC$. The different colours represent the different statistics used when fitting the TNG300 clustering, as labelled. The probability distribution function for each parameter is displayed at the top of each column. The dashed and solid lines represent the credibility regions derived from clustering statistics with and without the lensing statistic.}  
\label{Fig:mcmc}
\end{figure*} 

In this section, we will demonstrate that \shame is capable of simultaneously fitting the clustering and lensing statistics of various TNG300 galaxy samples. For the fitting, we used the same parameters space as the one used to train the emulator, assuming flat priors.

As we will show in \S~\ref{sec:baryons}, baryonic effects can have a significant impact on scales below $0.6~\hMpc$; therefore, we will only fit the GC and GGL to scales above that. For completeness, we will also present results on scales smaller than $r=0.6~\hMpc$. 

To find the best-fitting \shame parameters, we  minimise the $\chi^2$, which we computed as:

\begin{equation}
\label{eq:chi2}
\chi^2 = (V_{\rm TNG300}-V_{\rm SHAMe})^T C_v^{-1} (V_{\rm TNG300}-V_{\rm SHAMe}),
\end{equation}

\noindent with $V_{\rm TNG300}$ and $V_{\rm SHAMe}$ the concatenated vector of our summary observables for the TNG300 and for  the SHAMe mocks respectively, and $\rm C_v^{-1}$ the inverse of the covariance matrix.

We sample the posterior distribution using the Monte-Carlo Markov Chain (MCMC) algorithm, which is implemented in the public code {\tt emcee} \citep{emcee}. The likelihood function was computed as $2 \ln\ \mathcal{L} = -\chi^2$, with $\chi^2$ computed using Eq.~\ref{eq:chi2}. We utilised 1,000 chains with individual lengths of 10,000 steps and a 1,000-step burn-in phase. Although uncommon, this combination of chains and steps is ideal for an emulator-based MCMC, which is extremely efficient when computing a large number of points simultaneously. Additional MCMC configurations were tested, with nearly identical outcomes. Each MCMC chain takes an average of $\sim 20$ minutes to compute on a single CPU.

We show the results of \shame with the best fitting parameters in  Fig.~\ref{Fig:gal_cluster}. Each panel shows the results for one of our 4 statistics, as specified in the plot. Black solid lines show the measurements in the TNG300 simulations, whereas the red lines show our best-fitting \shame result. Within each panel, bottom, middle, and top sets of lines display the result for three samples with number densities equal to  $10^{-2},\ 3.16\times10^{-3}\ \&\ 8.7\times10^{-4}\ \ihMpcC$, respectively. For the sake of clarity, the lowest number density has been shifted in the plot. 

The \shame model performs well for all number densities and clustering statistics. There are minor differences in the lensing at small scales, which is expected as baryonic effects become increasingly significant in this regime. It's worth noting that the TNG300 is a hydrodynamic simulation with 100 times the mass resolution of our dark matter simulation. This simulation replicates a wide range of observables while following complex astrophysical processes, modelling the baryon-dark matter interaction. The ability of our model to reproduce galaxy clustering in a matter of seconds with only five free parameters is impressive, but the ability to recover the galaxy lensing signal in the absence of baryonic effect up to scales of $0.6 \hMpc$ demonstrates the model's robustness. Furthermore, in appendix~\ref{sec:rmin}, we show the fits when all scales larger than $0.1\hMpc$ are icluded, finding that our model can reproduce the GC (but not the GGL) over the full range of scales.

The excellent agreement in these joint constraints suggests that for these scales and samples, the \shame model does not exhibit any systematic discrepancy between the correlation functions and the lensing signal, such as the one expected by the so-called ``Lensing-is-low" problem \citep {Leauthaud:2017}. This effect is an overestimation of the lensing signal for some galaxy samples when the projected correlation function is fitted using a HOD method. Nonetheless, this only suggests, not proves, that our model is free of the ``Lensing is low" effect. A detailed study on the origin of this problem can be found in \cite{ChavesMontero:2022}. 

In Fig. \ref{Fig:mcmc} we present the constraints on the free parameters of \shame obtained by employing different combinations of GC and GGL data for fitting the TNG300 in our intermediate density sample, $\bar{n}=3.16\times10^{-3}\ihMpcC$. The lines represent $1\,\sigma$ confidence regions, colour-coded by the different statistics used when fitting, as labelled.

Constraints using only the projected correlation function and the GGL are shown in black and purple lines, respectively. Within the statistical uncertainty, we can see that the posteriors in all \shame parameters are compatible. In the case of $\sigL$ and $\tmerger$, they seem to provide roughly identical constraining power, whereas for the assembly bias parameters, $\FkP$ and $\FkM$, these datasets become complementary. This can be better appreciated by the red contours, which show the results of simultaneously fitting \proj~and \lensing. It is evident that the posteriors get narrower and thus \shame parameters are better constrained. However, when redshift-space multipoles are included, the gain becomes much greater, as shown by the blue lines. We can see better-constrained parameters in this case, particularly for the galaxy merger timescale. Furthermore, as shown by the cyan lines, including GGL contributes relatively little to the posteriors measured by GC statistics. We verified that all of these results hold qualitatively when larger minimum scales are used. We also find similar results for other galaxy number densities, with the lensing signal becoming slightly more important for the lowest number density, but still with a negligible contribution when compared to the multipoles of the correlation function.

\begin{figure*}
\includegraphics[width=1.0\textwidth]{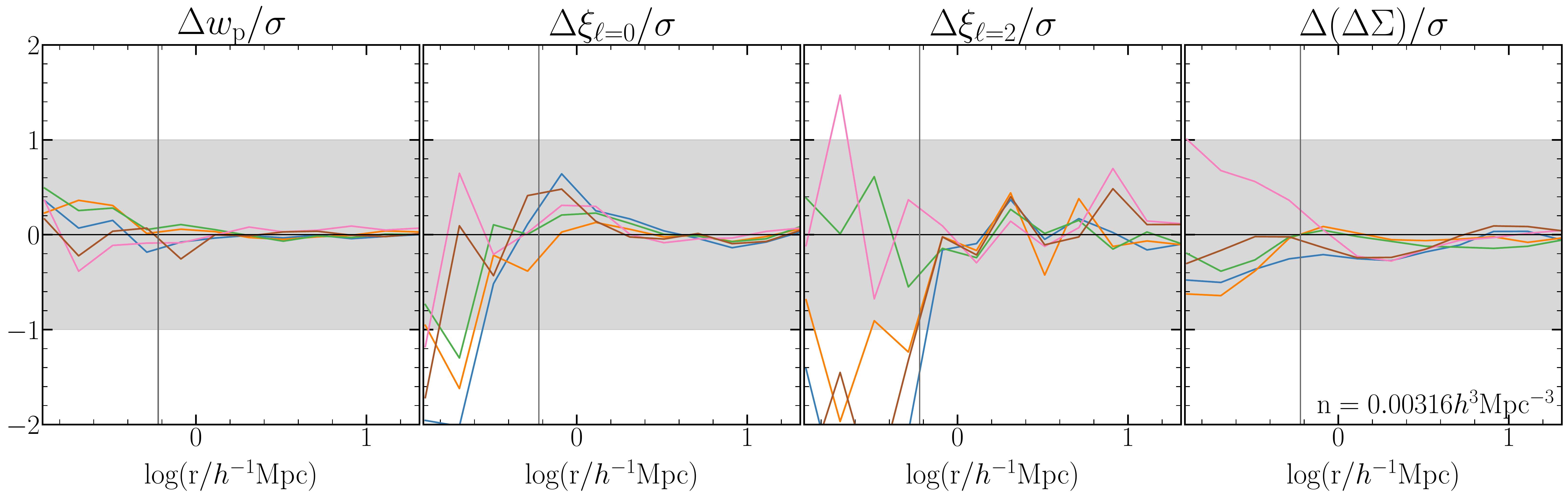}
\caption{The ratio between the galaxy clustering statistics (projected correlation function, monopole, quadrupole and lensing) of five different SAMs and those of a SHAMe model fitted to each SAM for a galaxy number density of $n=3.16\times10^{-3}\ \ihMpcC$. The fit is done for scales above 0.6 $\hMpc$, denoted as a vertical line on the figures. Other number densities show similar results. As described in \S~\ref{sec:cluster}, the \texttt{L-Galaxies} SAMs have different astrophysical parameters, leading to distinct clustering predictions.} 
\label{Fig:SAM}
\end{figure*}


To further validate our method, we compared its galaxy clustering predictions to those of five semi-analytical models of galaxy formation (SAMs, see \S~\ref{sec:sam} for more details). In Fig.~\ref{Fig:SAM}, we show the differences between the GC and GGL of the SAM and the \shame model. It is evident from the plot that the \shame model can indeed reproduce all four clustering statistics for each of the five SAM models.

In contrast to hydrodynamic simulations, the baryonic matter is tracked analytically in SAMs via the merger trees of the simulation and therefore has no effect on the distribution of dark matter. Thus, \shame has no difficulty reproducing GGL in these models, even at small scales. As a result, we did not select SAMs as the primary test of the \shame model, but they are crucial for demonstrating that GC and GGL can be reproduced simultaneously for a wide variety of galaxy formation models.  For consistency with the TNG300 measurements, we only fit the GC and GGL up to $0.6~\hMpc$; however, the \shame model can successfully reproduce these statistics at least up to $0.1~\hMpc$ (see appendix~\ref{sec:rmin} for more details.)

\section{The effects of baryons}
\label{sec:baryons}

An important feature of hydrodynamic simulations, such as the TNG300, is that they directly follow the response of the dark matter distribution to astrophysically driven changes in the distribution of baryons. While we do not anticipate that baryon-dark matter interactions will alter the mass distribution on large scales, there may be significant changes on non-linear scales, specifically in the 1-halo term regime. In particular, baryonic physics, such as AGN feedback, supernova feedback, and star formation, will alter the matter distribution compared to the predictions of a gravity-only simulation. Naturally, these effects are absent from our \shame predictions, which may restrict the scales at which our mock galaxy catalogues unbiased.

To determine how much of the difference between the GGL of the \shame model and the TNG is due to baryonic effects, we displaced the dark matter particles of the TNG300-mimic simulation to replicate the expected mass distribution of the TNG300 hydrodynamic simulation using the baryonification technique described in \cite{Arico:2020,Arico:2021a}. This method shifts the dark matter particles of a gravity-only simulation to account for the distribution of stars and gas (ejected and bound) within a dark matter halo, as well as the central galaxy. Aric{\`o} et al. used a 7-parameter baryonification model to fit the power spectrum and bispectrum of the hydrodynamic simulation up to $k=5\ihMpc$ in order to determine the model that most accurately reproduces the baryonic effects of a particular hydrodynamic simulation (in this case, the TNG300).

\begin{figure}
\includegraphics[width=0.45\textwidth]{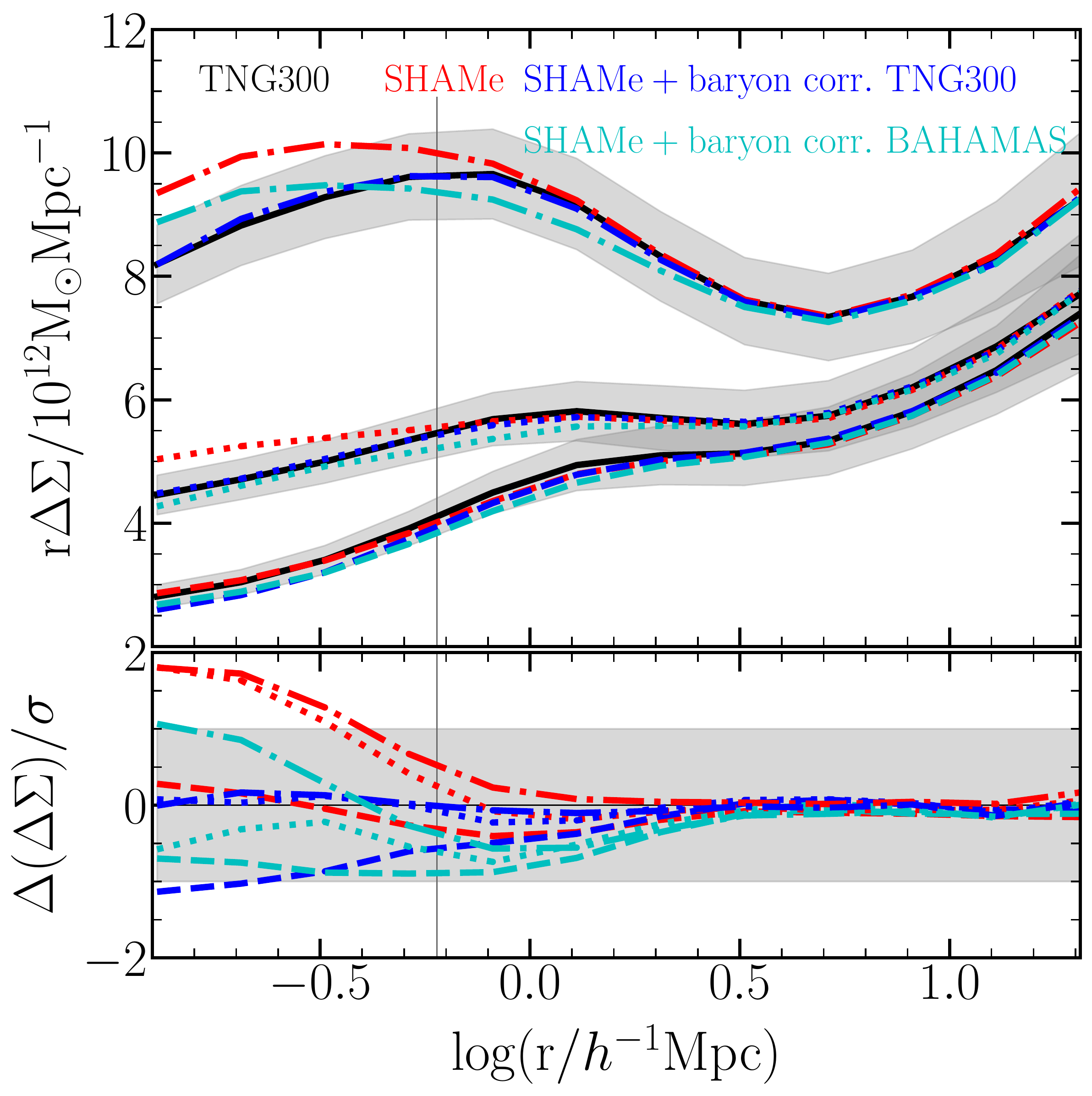}
\caption{The galaxy lensing signal (\lensing) for the TNG300 (black lines), the SHAMe model (red lines) and the SHAMe model with a baryon correction model, using the best fitting parameters for the TNG300 (blue lines) and for the BAHAMAS simulations (cyan lines).}  
\label{Fig:baryons}
\end{figure}

In Fig.~\ref{Fig:baryons}, we show the GGL over a dark matter simulation with baryonic effects using a SHAMe with the best-fitting parameters previously found by our emulator. We find a significant improvement in our two least dense samples when we use the baryonification model that best fits the TNG300, being able to reproduce the GGL signal down to $r=0.1~\hMpc$. We find a slight decrease in fit accuracy for the densest sample, but it is not as significant as the improvement found for samples with lower densities. This is likely due to the baryonification model determining the optimal parameters by analysing the simulation's entire power spectrum. We do not expect the baryonic effect caused by low halo masses to have a significant impact on this statistic, so the constraint on the baryonic effects in these haloes is lower. In fact, we have checked that the baryon fractions in low-mass haloes are significantly underpredicted by our baryonification model. Furthermore, \cite{Arico:2021a}'s fit of the baryonic effects does not focus on scales as small as the one shown here, reducing the precision in this regime. Despite these limitations, we find that the overall agreement with the lensing signal of all our samples is excellent, indicating that the main limitation of \shame in reproducing the TNG300 lensing signal is due to the lack of baryonic effects in the dark matter simulation to which it is applied.

In addition to the TNG300-baryonification model, we examine how much the GGL changes if we implement a baryonification model designed to match a different hydrodynamical simulation, specifically the BAHAMAS simulation \citep{McCarthy:2017,McCarthy:2018}. In Fig.~\ref{Fig:baryons}, the predicted GGL when the same \shame  model is applied to the same simulation modified by this different baryonification model is shown in cyan. In contrast to the TNG300 case, where we only detect baryonification effects on scales below $0.6~\hMpc$, the change in the GGL signal is present on scales below $1~\hMpc$ and is significantly different from the other GGL calculations.

These variations in GGL suggest two possibilities. If we wish to remain agnostic about baryonic effects on our target samples (as is the case when working with observations) and if our modelling is based on dark-matter-only simulations, then the GGL signal on small scales ($<1~\hMpc$) should not be fitted without a flexible baryonification model or a covariance matrix that accounts for the expected error on these scales. Ignoring this may result in systematic errors in the predictions as a result of improper small-scale fits. Because the differences produced by baryonic effects on GGL are significant, it may be possible to use GGL to constrain baryonic effects by combining SHAMe with a baryonification model such as \cite{Arico:2021a}. This will be addressed in a future paper.

\section{Other statistics}
\label{sec:other_stat}
In the previous section, we demonstrated that the \shame model can simultaneously reproduce the GC and GGL of the TNG300 and semi-analytic models run with various physical prescriptions. Only five free parameters were required to fit \proj,~\mono,~\quadr~\&~\lensing~ for scales above 0.6 $\hMpc$. To further validate the \shame model, we will evaluate its performance when reproducing statistics not used when fitting the \shame parameters. In \S~\ref{sec:richness}, we show the predicted halo occupation distribution and halo richness, and in \S~\ref{sec:gab} we will look into galaxy assembly bias predictions of the \shame model.

\subsection{Group richness}
\label{sec:richness}

The halo occupation distribution (HOD) was originally developed for the purpose of interpreting galaxy clustering. This distribution specifies the average number of galaxies in a halo as a function of halo mass. Ignoring any galaxy assembly bias effects, two different models with the same HOD and running on the same dark matter simulation should have the same large-scale bias. It is important to note that the inverse is not necessarily true: having identical large-scale clustering does not ensure that HODs are identical. Furthermore, even with the same HOD, the distribution of mass and galaxies within dark matter haloes could be different, meaning there could be differences in the GC and GGL on small scales \citep{ChavesMontero:2022}.   

\begin{figure*}
\includegraphics[width=0.45\textwidth]{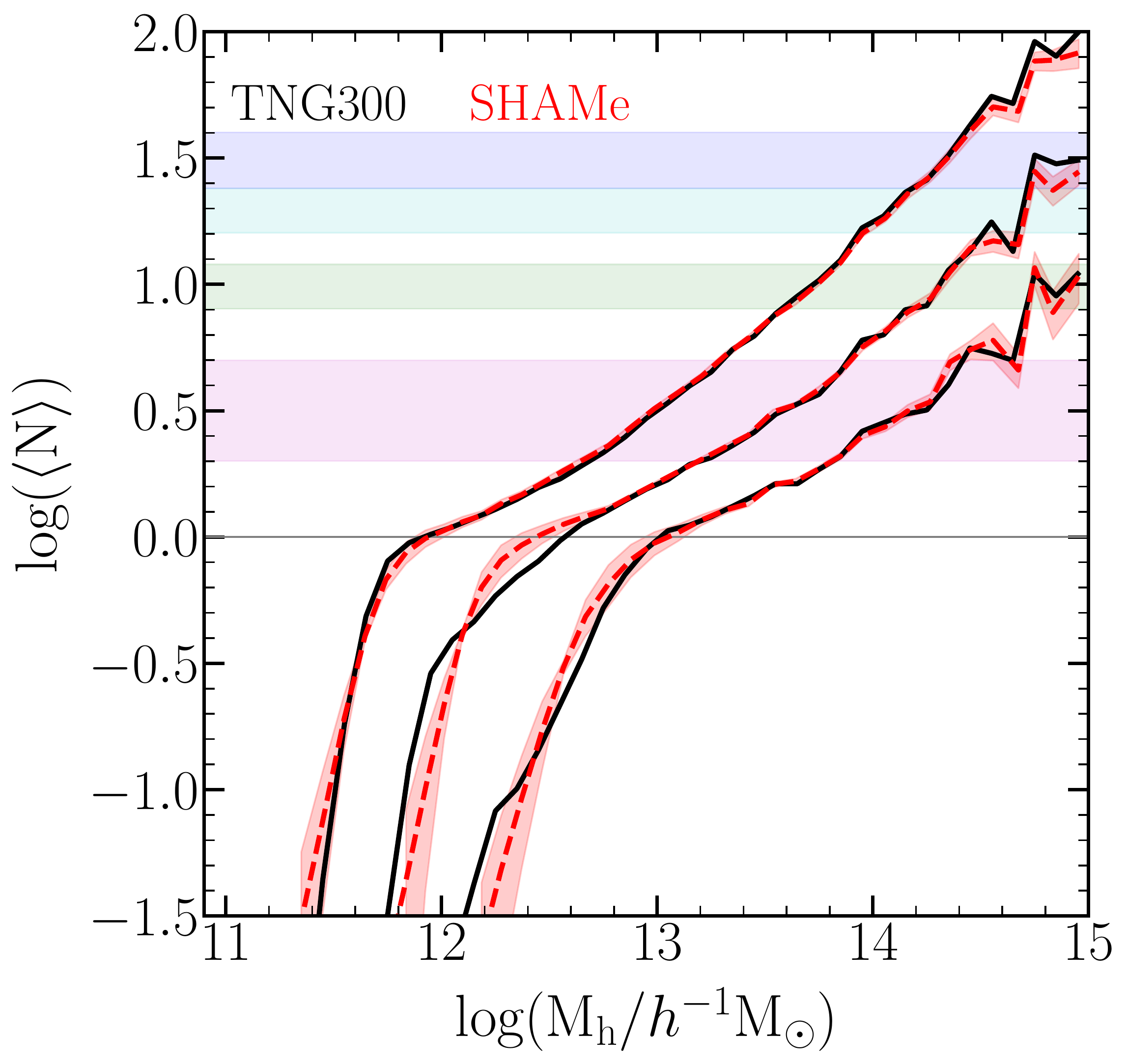}
\includegraphics[width=0.45\textwidth]{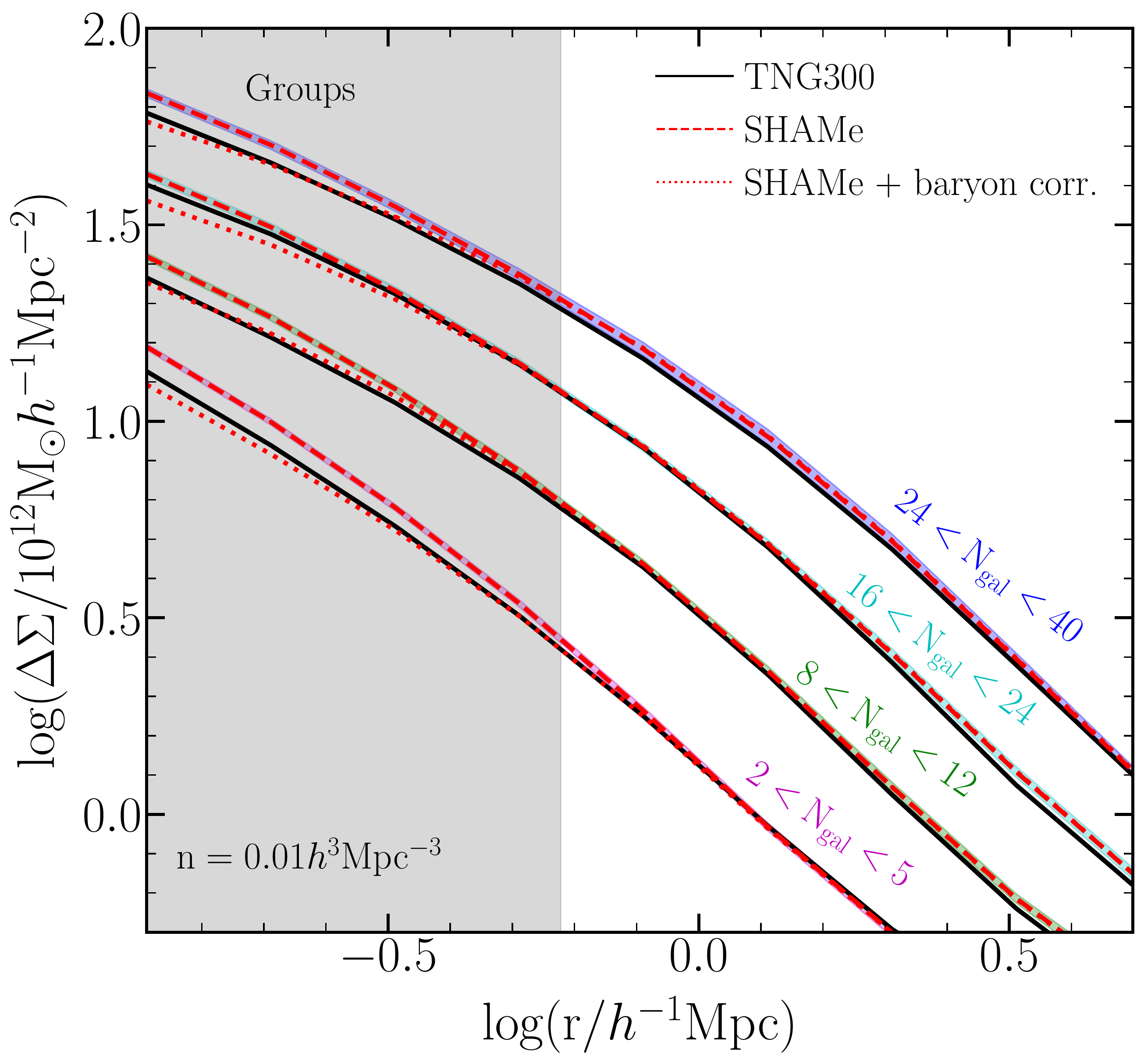} 
\caption{(Left panel) The halo occupation distribution (HOD), which measures the average number of galaxies occupying haloes as a function of their masses. The black lines represent the predictions of the TNG300, whereas the red lines and shaded regions represent the mean and standard deviation of  200 different SHAMe mocks with parameters chosen randomly from the MCMC used to fit the TNG300's GC and GGL (see \S~\ref{sec:cluster} for more details). The three curves are for galaxy samples with luminosity cuts corresponding to number densities of $0.01,\ 0.00316\ \&\ 0.00087\ \ihMpcC$. (Right panel) Stacked weak lensing profiles for the TNG300 and the SHAMe mocks around haloes that contain [2,5], [8,12], [12,24] and [24,40] galaxies from the largest number density sample ($0.01\ \ihMpcC$). The typical masses of these halos can be inferred from the coloured bands in the left panel. The coloured shaded region corresponds to one standard deviation around the mean for 100 different SHAMe mocks. The shaded region represents the scales not used by the SHAMe to fit the TNG300 clustering. For clarity, the lines have been shifted along the y-axis by 0.3 dex.} 
\label{Fig:hod_Rich}
\end{figure*}

In Fig.~\ref{Fig:hod_Rich} we show HODs for the TNG300 simulation and the SHAMe catalogues. The solid lines and shaded regions for the SHAMe models represent the mean and one sigma dispersion of 200 SHAMe catalogues with parameters corresponding to random points in the MCMC when \proj, \mono, \quadr and \lensing~ were used as constraints. 
To account for differences in halo mass function between the TNG300 hydrodynamic simulation and the dark-matter-only TNG300-mimic simulation, we match the halo masses between the simulations for the TNG300-mimic to reproduce exactly the TNG300 mass function. This is a common approach when comparing HODs with different halo mass functions (e.g. \citealt{C13,C17}). We do this by (i) computing the difference between the cumulative halo mass function of the hydrodynamic and dark-matter-only simulation. (ii) expressing this difference as a function of the halo mass number density, and (iii) modifying the mass of the halos from the dark-matter-only simulation by this difference.
We then find good agreement between the original and the \shame HOD. In particular, in the satellite-dominated abundance regime, we find almost perfect agreement over all halo masses.

We now look to the lensing signal that comes from haloes with similar group richness. We explore this by measuring the stacked lensing signal around groups with occupation numbers in the ranges $2<{\rm N_{gal}< 5}$, $8<{\rm N_{gal}< 12}$, $16<{\rm N_{gal}< 24}$ and $24<{\rm N_{gal}< 40}$ for a number density of $n=0.01 \ihMpcC$ (indicated as horizontal coloured bands in the left panel of Fig.~\ref{Fig:hod_Rich}). We find reasonably good agreement with the TNG300, especially after including baryonic effects. This test highlights the capacity of the \shame model to populate dark matter haloes similarly to the hydrodynamic simulation, even when the halo occupation distribution is not directly used to fit the model. We also tested a number density of 0.00316 $\ihMpcC$, finding similar results.

\subsection{Galaxy assembly bias constraints}
\label{sec:gab}

Another statistic we can infer from galaxy clustering that is not directly fitted by the \shame model is galaxy assembly bias \citep{Croton:2007}. This is a change in large-scale galaxy clustering that can be induced by the known dependence of halo clustering on halo properties other than mass (e.g. halo assembly bias, \citealt{Gao:2005,Gao:2007}). If the halo occupation of a galaxy sample also depends on these secondary halo properties (sometimes known as occupancy variation, \citealt{Zehavi:2018, Artale:2018}), then large-scale galaxy clustering changes. For various galaxy formation models, target number densities, and target redshifts, the amplitude of the galaxy assembly bias signal can vary, but previous studies (e.g. \citealt{ChavesMontero:2016,C19,C21a}) have estimated an impact between -20\% and 20\% on the two-point correlation function signal.

Due to the fact that our model has a variable amount of assembly bias, we can constrain galaxy assembly bias in our target sample based on its GC and GGL signals. To accomplish this, we select 100 random points along the MCMC used to fit the galaxy clustering signals, use these parameters to build new mocks (not with the emulator, but rather by directly populating the dark matter simulation), and calculate their assembly bias level using the shuffling technique developed by \cite{Croton:2007}. This method involves exchanging populations of galaxies between haloes of similar mass (in bins of 0.1 dex in ${\rm log}(\hMsun))$ while maintaining the relative distribution of galaxies within a halo. This shuffle run will, by definition, have the same HOD and 1-halo clustering signal as the original sample. The galaxy assembly bias of the original sample is then defined as the square root of the ratio of the two-point correlation function of the original run to that of the shuffled run:
\begin{equation}
\xi/\xi_{\rm shuffle} = b^2.
\end{equation}
To reduce the noise of this ratio, we shuffled each of the 100 mocks five times and the TNG300 twenty times. 

\begin{figure}
\includegraphics[width=0.45\textwidth]{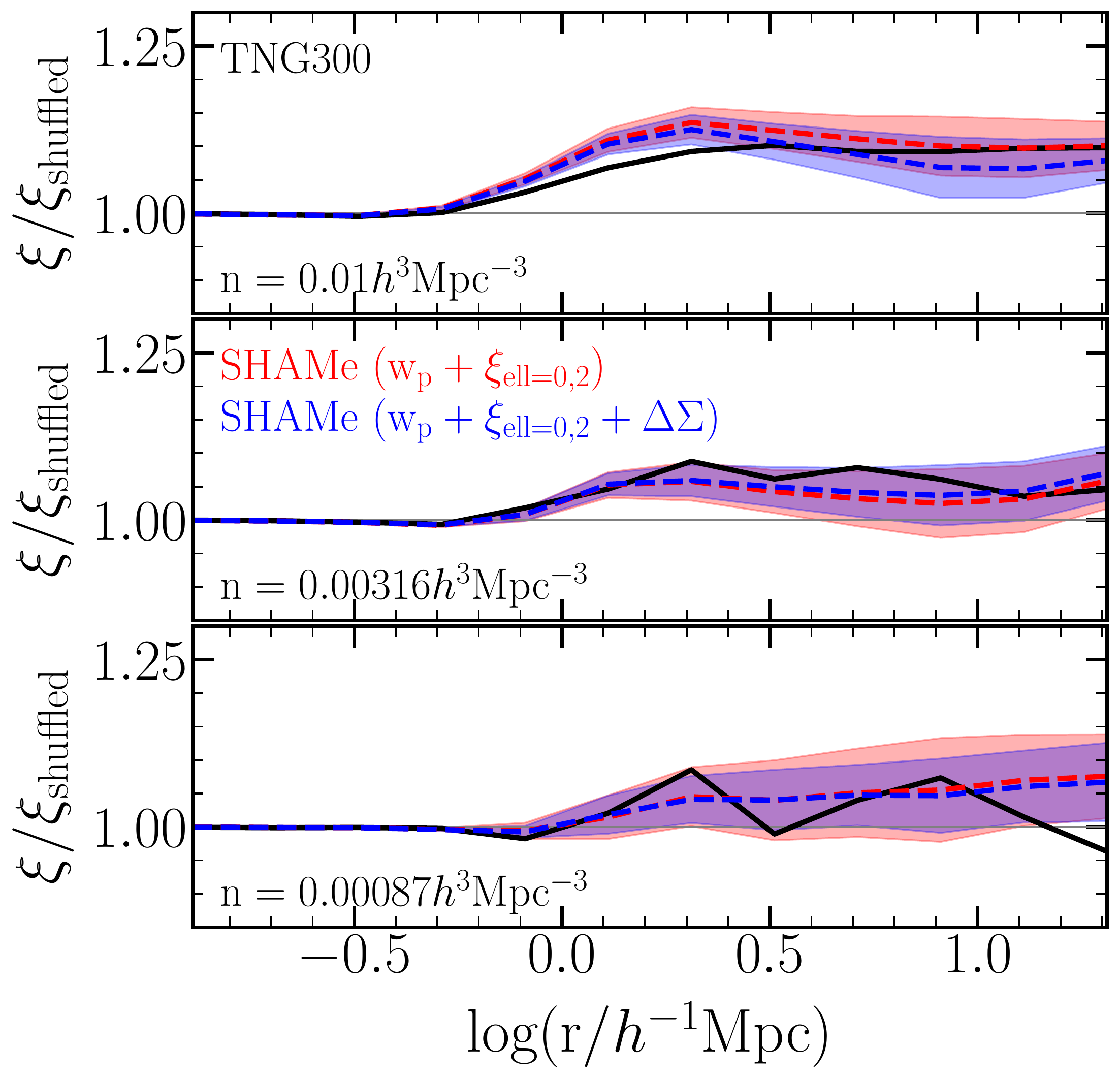}
\includegraphics[width=0.45\textwidth]{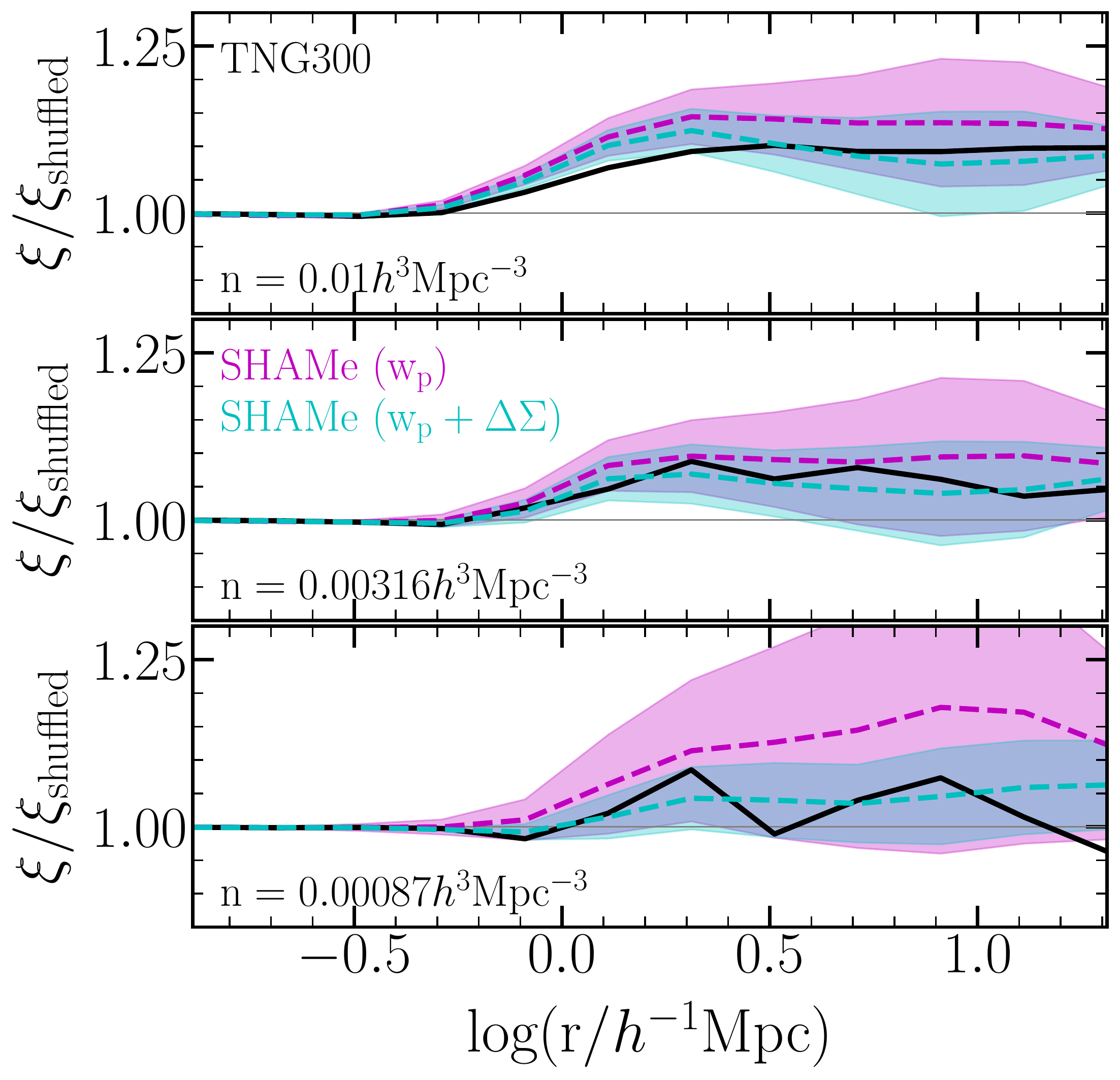}

\caption{Real-space correlation functions of the TNG300 (black line) and multiple \shame models (coloured lines) divided by their shuffled counterparts. The black lines represent the ratio using an average of 20 shuffled runs for the TNG300. For the \shame mocks, the shaded region corresponds to one standard deviation around the mean for 100 different mocks taken from the MCMC, each of them divided by five shuffled runs. The top panel compares the TNG300 galaxy assembly bias estimates to \shame mocks that were fitted to reproduce \proj, \mono ~\& \quadr and \proj, \mono, \quadr ~\& \lensing, whereas the bottom panels compare the TNG300 estimates to \shame models that were fitted to reproduce only \proj and \proj~\& \lensing. The galaxy assembly bias of the TNG300 is correctly recovered for all number densities and for all combinations of statistics within one sigma.}  
\label{Fig:gab}
\end{figure}

The predicted galaxy assembly bias signal of the TNG300 is compared to that of the \shame mocks in Fig.~\ref{Fig:gab}. In the upper panel, we compare the predictions obtained when fitting \proj,~\mono~\&~\quadr with the predictions obtained when fitting the same functions plus \lensing. We find that in both cases and for all galaxy number densities, \shame estimates the assembly bias to within one sigma. Incorporating the lensing signal provides no additional constraints for the highest number densities (the faintest galaxies), and a marginal improvement on the constraints for the lowest number density. In the bottom panel we compared the galaxy assembly bias computed when fitting just \proj and \proj ~\&~\lensing. Again, the correct level of assembly bias is constrained within one sigma at all cases. However, when GGL is included, the constraints improve significantly. This section leads us to conclude that, for spectroscopic surveys, the multipoles of the correlation function will be sufficient to constrain galaxy assembly bias, with little constraining power added by GGL. However, if the multipoles are unavailable, as in surveys with photometric redshifts, the  lensing can improve constraints on galaxy assembly bias by up to a factor of two. It is important to note that these conclusions only apply to \shame-like models. 

Contrary to \cite{C22a}, we find that galaxy assembly bias can be constrained even for the lowest number density, whereas they were unable to recover the correct assembly bias level for those luminosity cuts. This difference could have several origins. Here we use separate assembly bias parameters for central and satellite galaxies, whereas \cite{C22a} forced the two parameters to be equal. This was because a single parameter seemed sufficient for optimal cosmological constraints, the primary goal of \cite{C22a}. Here, however, we focus on maximising information about the galaxy formation model, which leads us to vary the two parameters independently. Another possibility could be that here we apply the \shame model to a dark matter simulation with the same (known) cosmology as the TNG300, while \cite{C22a} marginalized over cosmology when obtaining their constraints. Finally, the simulations used in this work (the TNG300) are different from those used by \cite{C22a} (the Millennium-TNG simulation). We plan to look into this in a future paper.

\section{Summary} 
\label{sec:summary}

In this paper, we examine how well a dark-matter-only simulation with haloes populated using the subhalo abundance matching extended (\shame) model, a physically motivated empirical model, can reproduce the redshift-space clustering and the galaxy-galaxy lensing predicted by a hydrodynamic simulation, the TNG300. The \shame model \citep{C21a,C21c} improves upon the earlier SHAM model by incorporating an orphan prescription, (allowing us to reproduce galaxy samples with high number densities even within low-resolution simulations) an attenuation of the luminosity of satellite galaxies (to account for effects such as stripping and tidal disruption) and a variable amount of galaxy assembly bias.

We use the SHAMe model to create an emulator that can reproduce the GC and GGL signals in a fraction of a second for any chosen set of parameters. We use this emulator to fit clustering and lensing in the TNG300. Using MCMC, we explore the additional constraints that the lensing signal can offer. Here, we highlight our key findings:

\begin{itemize}

\item Using the \shame model, we can reproduce the projected correlation function, the monopole and the quadrupole of the TNG300 galaxy distribution down to scales of 0.1 $\hMpc$, and its lensing signal down to scales of 0.6 $\hMpc$ (Fig.~\ref{Fig:gal_cluster}). Baryonic effects prevent us from matching the lensing signal at smaller radii.

\item In addition, we were able to simultaneously fit GC and GGL in several semi-analytical galaxy formation models over the full range $ 0.6 \hMpc<r<20 \hMpc$ (Fig.~\ref{Fig:SAM}). These models have differing astrophysical prescriptions, demonstrating that our methodology works regardless of the characteristics of our target sample. 

\item By examining the marginalised credibility regions produced by the MCMCs, we can infer that the lensing signal does not add constraining power beyond that of the multipoles of the correlation function. However, it can significantly tighten constraints based on the projected correlation function alone.  (Fig.~\ref{Fig:mcmc}).

\item Modifying the dark matter simulation to which SHAMe is applied using a model for baryonic effects tuned to match TNG300 allows the TNG300 lensing signal to be matched within one sigma down to 0.1 $\hMpc$ (Fig.~\ref{Fig:baryons}). We show that the lensing signal is sensitive to the particular baryonification model used on scales below about 1 $\hMpc$.

\item We compute halo occupation distributions (HODs) for SHAMe mocks that are good fits to GC and GGL in the TNG300, finding excellent agreement both among the models and with the HOD of TNG300 itself. Thus, for a given halo mass, the mean number of galaxies per halo is very similar in the mocks and in the TNG300. The stacked lensing signal of haloes with a similar occupation is also similar among the models, showing how accurately the \shame model can populate the haloes of a hydrodynamic simulation (Fig.~\ref{Fig:hod_Rich}).

\item When fitting the TNG300's GC and GGL, the SHAMe model  reproduces the galaxy assembly bias signal of the TNG300. When the multipoles of the correlation function are used, the lensing signal does not significantly improve the constraints on assembly bias, but it significantly strengthens such constraints when only the projected correlation function is available (Fig.~\ref{Fig:gab}).

\end{itemize}

We conclude that the \shame model can reproduce the galaxy-galaxy lensing signal of complex models such as hydrodynamic simulations or semi-analytic galaxy formation models simultaneously with their galaxy clustering signal for $r$-band samples. In situations where precise redshifts are not available for galaxies (e.g., photometric surveys), including the lensing signal can significantly strengthen parameter constraints. However, if the multipoles of the redshift-space correlation functions are available, the addition of the lensing signal provides little improvement on the constraints when assuming fixed cosmology. 

Regarding baryonic effects, if their influence on the target sample is greater than in the TNG300, they may affect the lensing signal on scales larger than the smallest scale used in this study (0.6 $\hMpc$). Combining \shame with an appropriate baryonification model allowed us to fit down to smaller scales, and it may therefore allow marginalisation over uncertain baryonic physics. We plan to study this in a future paper.

\section*{Data availability}
The IllustrisTNG simulations, including TNG300, are publicly available and accessible at \url{www.tng-project.org/data} \citep{Nelson:2019}. The data underlying this article will be shared on reasonable request to the corresponding author.

\section*{Acknowledgements}

We thank Daniele Spinoso for the assistance when running l-galaxies.  
SC acknowledges the support of the ``Juan de la Cierva Incorporac\'ion'' fellowship (IJC2020-045705-I).
REA and JCM acknowledge support from the ERC-StG number 716151 (BACCO). 
REA from the Project of excellence Prometeo/2020/085 from the Conselleria d'Innovaci\'o, Universitats, Ci\`encia i Societat Digital de la Generalitat Valenciana and JCM from the European Union's Horizon Europe research and innovation programme (COSMO-LYA, grant agreement 101044612). IFAE is partially funded by the CERCA program of the Generalitat de Catalunya.
The authors also acknowledge the computer resources at MareNostrum and the technical support provided by Barcelona Supercomputing Center (RES-AECT-2019-2-0012 \& RES-AECT-2020-3-0014)

\bibliographystyle{mnras}
\bibliography{Biblio}

\appendix

\section{Predictions for lower scales}
\label{sec:rmin}
\begin{figure*}
\includegraphics[width=0.95\textwidth]{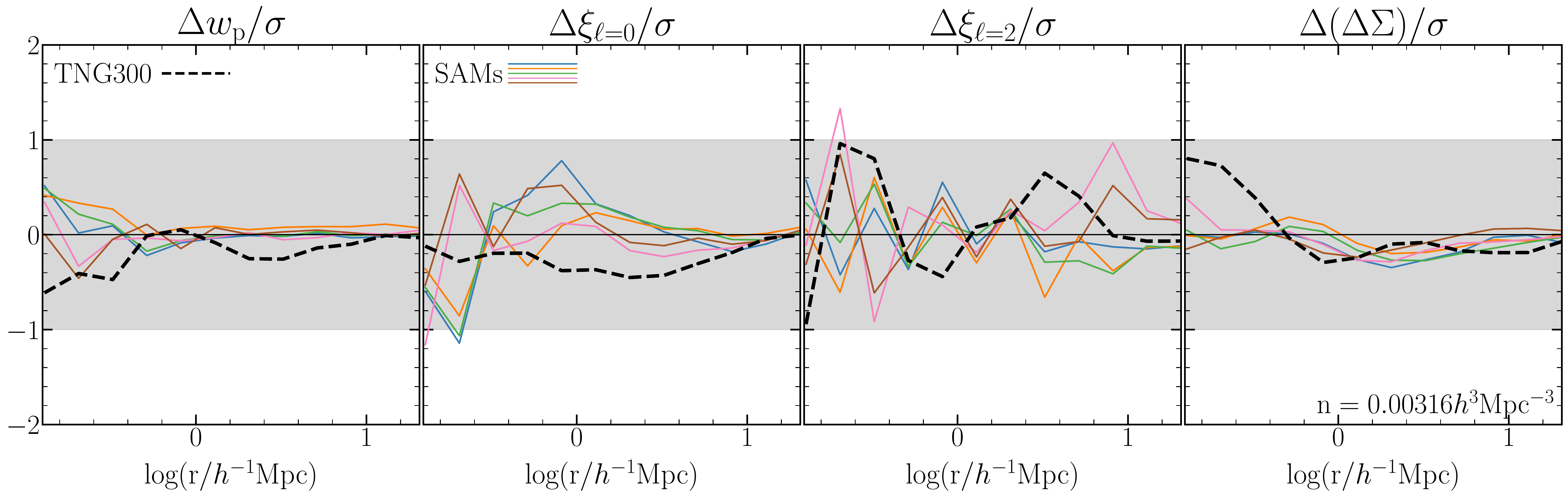}

\caption{Similar to Fig.~\ref{Fig:SAM}, but including the TNG300 predictions and fitting all the scales above $0.1\ \hMpc$.}  
\label{Fig:rmin}
\end{figure*}

Throughout this paper, we fitted the GC and GGL for scales greater than 0.6 $\hMpc$, where baryonic effects are not dominant (see \S~\ref{sec:baryons} for more details). However, this does not imply that the GC predictions cannot be reproduced under this scale. In Fig.~\ref{Fig:rmin}, we show the GC and GGL predictions of the \shame model when fitting \proj,~\mono,~\quadr~\&~\lensing~of the TNG300 and the SAMs at scales greater than $0.1\ \hMpc$ for a number density of $0.0316\ \ihMpcC$. Other number densities showed similar predictions. We find that we can successfully reproduce the GC and GGL of the SAMs at all scales simultaneously. This makes sense given that these models are unaffected by baryonic effects.

For the TNG300, we are able to successfully reproduce the GC (\proj,~\mono~\&~\quadr) at all scales, but, as predicted, we are unable to successfully reproduce the inner scales of the GGL signal. Even though the deviations of the GGL signal are within one sigma, these systematic deviations can potentially bias the overall fit.

\bsp	
\label{lastpage}
\end{document}